\documentclass[12pt, a4paper]{article}
\usepackage[T1]{fontenc} 
\usepackage[utf8]{inputenc} 
\usepackage{hyperref}
\usepackage{geometry} 
\usepackage{graphicx} 
\usepackage{caption}
\usepackage{subcaption}
\usepackage{amsmath,amsfonts}
\usepackage {color}
\usepackage{authblk}

\title{Joint Optimization of seismometer arrays for the cancellation of Newtonian noise from seismic body waves in the Einstein Telescope}

\author[$\dag$ $^{1}$]{Francesca Badaracco}
\author[$^{2}$]{Jan Harms}
\author[$^{1}$]{Luca Rei}
\affil[$^{1}$]{INFN – Sezione di Genova. Via Dodecaneso, 33 – 16146 – Genova}
\affil[$^{2}$]{Gran Sasso Science Institute - Viale F. Crispi, 7 67100 L'Aquila }
\affil[$\dag$]{francesca.badaracco@ge.infn.it}
\vspace{10pt}

\begin{document}
\maketitle
\begin{abstract}
    Seismic Newtonian noise is predicted to limit the sensitivity of the Einstein Telescope. It can be reduced with coherent noise cancellation techniques using data from seismometers. To achieve the best results, it is important to place the seismic sensors in optimal positions. A preliminary study on this topic was conducted for the Einstein Telescope (ET): it focused on the optimization of the seismic array for the cancellation of Newtonian noise at an isolated test mass. In this paper, we expand the study to include the nested shape of ET, i.e., four test masses of the low-frequency interferometers at each vertex of the detector. Results are investigated in function of the polarization content of the seismic field composed of body waves. The study also examines how performance can be affected by displacing the sensor array from its optimal position or by operating at frequencies other than those used for optimization.
\end{abstract}
\section{Introduction}
Third-generation gravitational wave (GW) detectors will have enhanced sensitivity compared to current GW detectors. The Einstein Telescope (ET), in particular, is designed to push the observation band down to about 3\,Hz \cite{ET2020}. At these frequencies, two major hindrances are seismic noise and Newtonian noise (NN). Seismic noise can be suppressed using suspension systems and/or active seismic isolation. Newtonian noise is closely linked to seismic noise as it is produced by the gravity fluctuations induced by seismic waves that cause a density variation in the elastic medium where they propagate \cite{Harms2019}. ET will be built underground to reduce the impact of seismic noise (and therefore NN), but a NN cancellation system will still be necessary to achieve the ET design sensitivity. This system will consist of a Wiener Filter (WF) that estimates the NN affecting ET using data collected by a seismic array \cite{Coughlin2014, Coughlin2016}. The array configuration must be optimized to maximize the cancellation capabilities of the system \cite{DHA2012,Coughlin2016,Badaracco2019,BaEA2020}. In total, ET has 24 test masses (TMs); half of which form the high-frequency interferometers where NN plays a minor role. So, when it comes to NN in ET, we can focus on the 12 TMs of the ET low-frequency interferometers \cite{ET2020}. 

The work presented in Ref. \cite{Badaracco2019} is a first attempt to understand the geometry and robustness of a seismic array for NN cancellation in underground environments. Since the seismic field is composed of both P-waves (compression) and S-waves (shear), the effectiveness of an array is reduced because correlations are affected by the presence of two types of seismic waves, each one with different propagation velocities. Ref. \cite{Badaracco2019} optimized for a single TM and assumed a homogeneous and isotropic seismic field. This type of optimization would work well only for an isolated end TM of an interferometer. However, the two input TMs of the interferometer arms are close to each other, and due to ET's triangular configuration, the input TMs of one interferometer are only a few 100\,m away from the end test masses of another interferometer (see Fig \ref{fig:ET_triangle}). To perform the optimization of the array configuration around the four (relevant) test masses of the ET detector at each vertex, correlations of NN between TMs must be taken into account, and the cost function is defined by the residuals after noise cancellation.

In the following sections, we will introduce the cost function and equations necessary to evaluate it, then present and discuss the optimization results. 

\section{Cost function choice}
\begin{figure}
    \centering
    \includegraphics[width=0.5\textwidth]{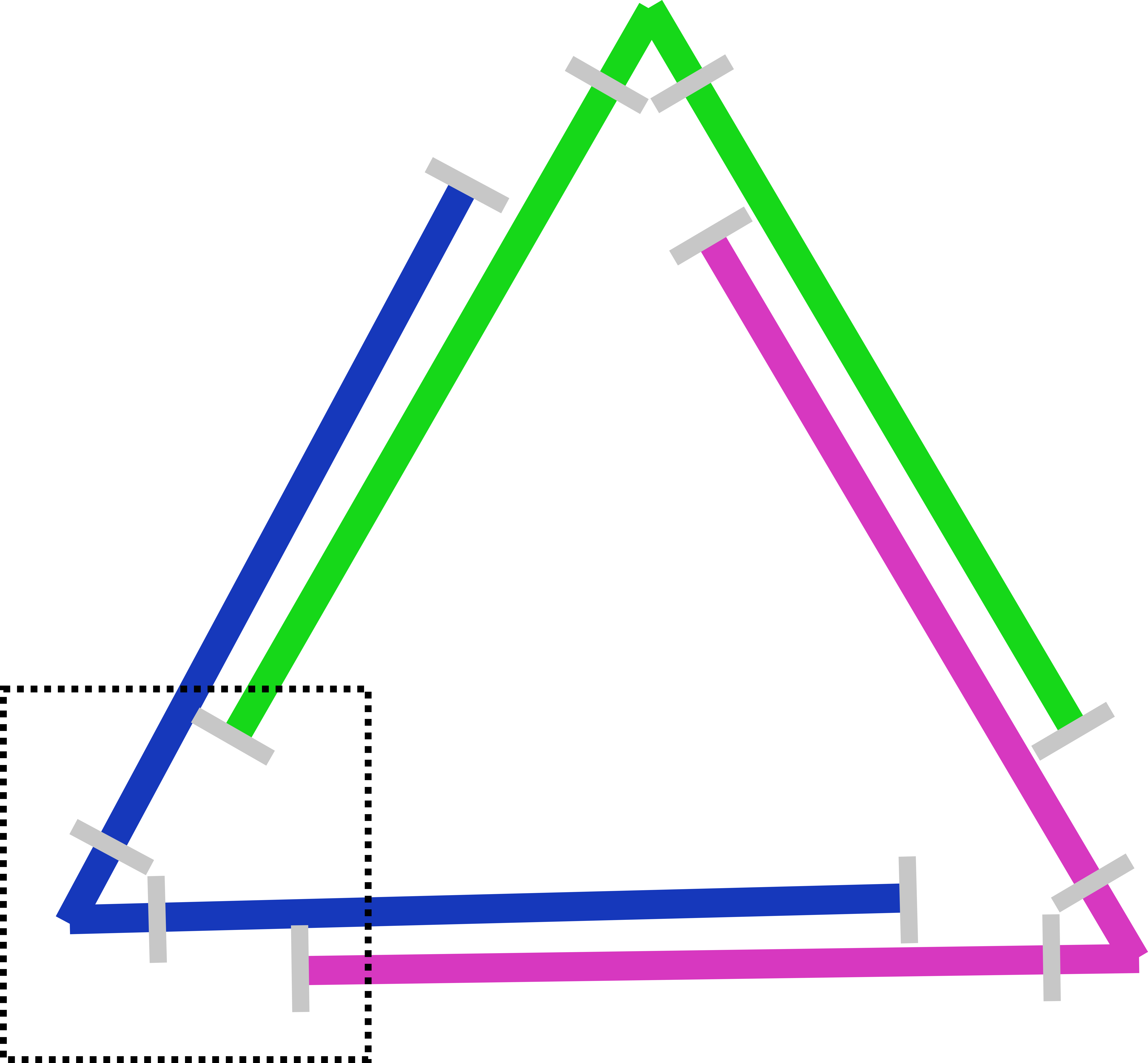}
    \caption{ET layout. At each vertex there are 4 test masses close to each other.}
    \label{fig:ET_triangle}
\end{figure}
An array optimization for ET should include all four TMs at a vertex to account for the possibility that a deployed seismometer can be used to subtract NN in all three ET Low-Frequency (ET-LF) interferometers (see black rectangle in Fig. \ref{fig:ET_triangle}). Optimizing arrays for each TM separately is less efficient (in terms of number of seismometers) than simultaneously considering all the TMs of one corner. Joint optimization can be performed using a cost function that takes into account NN from the four TMs and its correlations between TMs during each step of the optimization. The NN residuals left in all three ET-LF interferometers after subtracting out the estimated NN contributions from a vertex serve to define the cost function. The WF is built to be the optimal filter for linear problems such as NN cancellation \cite{Vaseghi2008}. The frequency domain residual can be expressed as in Ref \cite{Badaracco2019}: 
\begin{equation}\label{eq:residual}
    R(\omega) = 1 -\frac{\vec{C}_{\text{SN}}^\dag(\omega)\mathbf{C}_{\text{SS}}^{-1}(\omega)\vec{C}_{\text{SN}}(\omega)}{C_{\text{NN}}(\omega)},
\end{equation}
Where $\vec{C}_{\text{SN}}$ is the vector containing the cross-power spectral densities between the witness sensors (the seismic sensors in the array) and the target signal (ET-LF NN), $\mathbf{C}_{\text{SS}}$ is the matrix of the cross-power spectral densities between the witness sensors while $C_{\text{NN}}$ is the power spectral density of NN contributed by the TMs of a vertex. To optimize the array simultaneously considering the two input TMs of one interferometer and the end TMs of the other two interferometers, it is necessary to use both the residual for the two input TMs ($R_{\rm input}$) and two other residuals: one for each end TM of the other two interferometers ($R_{\rm end1}$ and $R_{\rm end2}$). The global cost function will then be defined as:
\begin{equation}\label{eq:optim}
\mathcal L  = \max_{\forall k \in ({\rm input, end1, end2}) } R_k(\omega)
\end{equation}
This type of joint optimization was also employed in Ref. \cite{Badaracco2019} to perform a frequency broadband optimization.
The meaning of Eq. \ref{eq:optim} is that at each optimization step, the algorithm attempts to optimize the worst residual between the three, thereby striving to obtain the best array for all the TMs.
This kind of cost function is convenient since we know that the residual relative to each interferometer will be equal or even less than the value indicated in the plots of the results (Section \ref{sec:results}). Moreover, we also tested an alternative cost function: the sum of the three residuals. Equation \ref{eq:optim} gives better single residuals (one for each interferometer) compared to the sum. The global cost function contains only three residuals instead of four because the two input TMs are accounted for in the same residual. Indeed, the correlation between a sensor close to the input TMs and the GW signal depends on both the input TM displacements. In the following Section it will be shown how to retrieve the residual for the input TMs.

\section{Residual for the input test masses}
Consider one vertex of ET. There will be two input TMs belonging to one interferometer and two end TMs belonging to the other two interferometers (see Fig. \ref{fig:ET_triangle}). In the following, we will focus solely on the two input TMs of one interferometer and calculate the cross-correlation between the target signal (represented by the input TMs) and one sensor. Assuming that the origin of the reference system is at one vertex of ET, then the two input TMs will be located at:
\begin{align}
\textbf{i}_1 =& r_{in}(1,0,0)\\
\textbf{i}_2 =& r_{in}(1/2, \sqrt{3}/2, 0)
\end{align}
With $r_{in}$ the distance of the input TM from the origin. The position variations of the end TM and input TM in one interferometer arm are denoted by $\delta x_{\rm e1}$ and $\delta x_{\rm i1}$, respectively. The total length of the interferometer arm is described by: $L_1 = L_0 + \delta x_{\rm e1} - \delta x_{\rm i1}$ for one arm and $L_2 = L_0 + \delta x_{\rm e2} - \delta x_{\rm i2}$ for the other arm, with $L_0=10$\,km. As a result, the signal detected by the interferometer will be: 
\begin{equation}
\delta L = L_2 - L_1 = (\delta x_{\rm e2} - \delta x_{\rm i2}) - (\delta x_{\rm e1} - \delta x_{\rm i1})
\end{equation}
However, the generic displacement $\delta x$ can also be written as a function of the NN acceleration: $\delta x = -\delta a_{\rm NN}/\omega^2$. We use an analytical NN model valid for spherical caverns sufficiently far underground (see \cite{HaEA2022} for a more detailed discussion of the underlying assumptions): 
\begin{equation}
\delta\textbf{a}_{\rm NN}(\textbf{r}_0,\omega) = 4\pi G\rho_0\left(2\boldsymbol{\xi}^{\rm  P}(\textbf{r}_0,\omega)\frac{j_1(k^{\rm P}r_c)}{k^{\rm P}r_c} - \boldsymbol{\xi}^{\rm S}(\textbf{r}_0,\omega)\frac{j_1(k^{\rm S}r_c)}{k^{\rm S}r_c}\right)
\end{equation}
where $r_c$ is the cavern radius and $\boldsymbol{\xi}^{\rm P,S}$ the seismic displacement of P and S seismic waves with wave vector $k^{\rm P,S}$, respectively. 
For $k^{\rm P,S}r_c\ll1$:
\begin{equation}\label{eq:NN_acceleration}
\delta\textbf{a}_{\rm NN}(\textbf{r}_0,\omega) = \frac{4}{3}\pi G\rho_0\left(2\boldsymbol{\xi}^{\rm P}(\textbf{r}_0,\omega) - \boldsymbol{\xi}^{\rm S}(\textbf{r}_0,\omega)\right)
\end{equation}
To calculate the residual (see Eq. \ref{eq:residual}), one needs the correlation between NN and seismic displacement measured at $\textbf{r}_s$. If the measurement axis of the sensor is along the unit vector $\hat{\textbf{e}}_s$, we find:
\begin{equation}
C_{\rm SN} = \left\langle \hat{\textbf{e}}_s \cdot \boldsymbol{\xi}(\textbf{r}_s,\omega),\delta L(\omega)/L_0\right\rangle.
\end{equation}
Seismic and NN correlations across the 10\,km arms are assumed to be zero in the NN band, therefore:
\begin{equation}\label{eq:Csn_target}
C_{\rm SN}(\omega) = \frac{1}{L_0\omega^2}\left\langle \hat{\textbf{e}}_s \cdot \boldsymbol{\xi}(\textbf{r}_s,\omega),(\delta a_{\rm i2}(\textbf{i}_2,\omega) - \delta a_{\rm i1}(\textbf{i}_1,\omega))\right\rangle
\end{equation}
Where, in the WF framework, $(\delta a_{\rm i2}(\textbf{i}_2,\omega) - \delta a_{\rm i1}(\textbf{i}_1,\omega))/(L_0\omega^2)$ will be considered the \textit{target signal}. Moreover, considering Eq. \ref{eq:NN_acceleration} and \ref{eq:Csn_target} we can write:
\begin{equation}
C_{\rm SN}(\omega) = \frac{1}{L_0\omega^2}\left\langle \hat{\textbf{e}}_s \cdot \boldsymbol{\xi}(\textbf{r}_s,\omega),\left(\hat{\textbf{e}}_2\cdot \delta\textbf{a}_{\rm NN}(\textbf{i}_2,\omega)-\hat{\textbf{e}}_1\cdot \delta\textbf{a}_{\rm NN}(\textbf{i}_1,\omega)\right)\right\rangle
\end{equation}
and then:
\begin{equation}
\begin{split}
C_{\rm SN}(\omega) &=  \frac{1}{L_0\omega^2}\frac{4}{3}\pi G\rho_0\\
&\quad\cdot\left\langle \hat{\textbf{e}}_s \cdot \boldsymbol{\xi}(\textbf{r}_s,\omega), \hat{\textbf{e}}_2\cdot\left(2\boldsymbol{\xi}^{\rm P}(\textbf{i}_2,\omega) - \boldsymbol{\xi}^{\rm S}(\textbf{i}_2,\omega)\right)-\hat{\textbf{e}}_1\cdot\left(2\boldsymbol{\xi}^{\rm P}(\textbf{i}_1,\omega) - \boldsymbol{\xi}^{\rm S}(\textbf{i}_1,\omega)\right)\right\rangle
\end{split}
\end{equation}
$C_{\rm SN}$ can be rewritten by defining $\boldsymbol{\xi}(\textbf{r}_s, \omega) = \boldsymbol{\xi}^{\rm P}(\textbf{r}_s, \omega) + \boldsymbol{\xi}^{\rm S}(\textbf{r}_s,\omega)$, and:
\begin{align}
\left\langle \hat{\textbf{e}}_s \cdot \boldsymbol{\xi}^{\rm P}(\textbf{r}_s,\omega), \hat{\textbf{e}}_k\cdot\boldsymbol{\xi}^{\rm P}(\textbf{i}_k,\omega) \right\rangle  &= \mathcal{C}(\xi^{\rm P}, \omega)f^{\rm P}(\Phi_{sk}^{\rm P})\\
\left\langle \hat{\textbf{e}}_s \cdot \boldsymbol{\xi}^{\rm S}(\textbf{r}_s,\omega), \hat{\textbf{e}}_k\cdot\boldsymbol{\xi}^{\rm S}(\textbf{i}_k,\omega) \right\rangle  &= \mathcal{C}(\xi^{\rm S}, \omega)f^{\rm S}(\Phi_{sk}^{\rm S})\\
\end{align}
where $\mathcal{C}(\xi^{\rm S,P}, \omega)$ represents the power spectral density of the seismic displacement $\xi^{\rm S,P}$ and $f^{\rm S,P}(\Phi_{sk}^{\rm S,P})$  is a function depending on the assumptions used for the seismic field model. In this case, a homogeneous and isotropic seismic field was assumed and $f^{\rm S}(\Phi_{sk}^{\rm S})$ and $f^{\rm P}(\Phi_{sk}^{\rm P})$ can be expressed as (see Ref. \cite{Harms2019}, Sec. 7.1.3):
\begin{align}
f^{\rm P}(\Phi^{P}_{\rm sk}) &= (j_0(\Phi^{P}_{\rm sk})+j_2(\Phi^{P}_{\rm sk}))(\hat{\boldsymbol{e}}_{\rm s}\cdot\hat{\boldsymbol{e}}_{\rm k}) - 3j_2(\Phi^{P}_{\rm sk})(\hat{\boldsymbol{e}}_{\rm s}\cdot\hat{\boldsymbol{e}}_{\rm sk})(\boldsymbol{e}_{\rm k}\cdot\hat{\boldsymbol{e}}_{\rm sk})\\
f^{\rm S}(\Phi^{S}_{\rm sk}) &= (j_0(\Phi^{S}_{\rm sk})-\frac{1}{2}j_2(\Phi^{S}_{\rm sk}))(\hat{\boldsymbol{e}}_{\rm s}\cdot\boldsymbol{e}_{\rm k}) + \frac{3}{2} j_2(\Phi^{S}_{\rm sk})(\hat{\boldsymbol{e}}_{\rm s}\cdot\hat{\boldsymbol{e}}_{\rm sk})(\hat{\boldsymbol{e}}_{\rm k}\cdot\hat{\boldsymbol{e}}_{\rm sk})
\end{align}
with: $\Phi_{sk}^{\rm P,S} = k^{\rm P,S}|\textbf{r}_s-\textbf{r}_k|$ and $\textbf{e}_{sk} = \left(\textbf{r}_s-\textbf{r}_k\right)/|\textbf{r}_s-\textbf{r}_k|$. Finally we get to the expression for $C_{\rm SN}(\omega)$:
\begin{equation}\label{eq:Csn}
C_{\rm SN}(\omega) = \frac{1}{L_0\omega^2}\frac{4}{3}\pi G\rho_0\,\mathcal{C}(\xi^{\rm tot}, \omega) \left[2p\left(f^{\rm P}(\Phi_{s2}^{\rm P}) -f^{\rm P}(\Phi_{s1}^{\rm P})\right) -(1-p)\left(f^{\rm S}(\Phi_{s2}^{\rm S}) -f^{\rm S}(\Phi_{s1}^{\rm S})\right)\right]
\end{equation}
where the indices $s$ and $1$-$2$ are to be interpreted as the indexes relative to the sensor and the two input TMs, while $p$ is defined as follows:
\begin{align}\label{eq:p}
p &= \frac{\mathcal{C}(\xi^{\rm P},\omega)}{\mathcal{C}(\xi^{\rm tot}, \omega)}\\
1-p &= \frac{\mathcal{C}(\xi^{\rm S},\omega)}{\mathcal{C}(\xi^{\rm tot}, \omega)}
\end{align}
Being $(\delta a_{\rm i2}(\textbf{i}_2,\omega) - \delta a_{\rm i1}(\textbf{i}_1,\omega))/(L_0\omega^2)$ the target signal, $C_{\rm NN}$ will read as:
\begin{equation}
C_{\rm NN} = \left\langle \left( (\delta a_{i2}(\textbf{i}_2,\omega) - \delta a_{i1}(\textbf{i}_1,\omega))/(L_0\omega^2) \right)^2 \right\rangle
\end{equation}
and following the same simple calculations as before, $C_{\rm NN}$ can be expressed as:
\begin{equation}\label{eq:Cnn}
C_{\rm NN} = \left( \frac{1}{L_0\omega^2}\frac{4}{3}\pi G\rho_0\right)^2 \mathcal{C}(\xi^{\rm tot}, \omega)\left( 2(3p + 1) -2\left( 4p f^{\rm P}(\Phi^{\rm P}_{12})+(1-p)f^{\rm S}(\Phi^{\rm S}_{12})\right)\right)
\end{equation}
with $\Phi_{12}^{\rm P,S} = k^{\rm P,S}|\textbf{i}_1-\textbf{i}_2|$. Finally, the correlation between P and S-waves have always been neglected.

Now, the residual (Eq. \ref{eq:residual}) for the input TMs of one interferometer can be evaluated using Eqs. \ref{eq:Csn}, and \ref{eq:Cnn} along with the value for $C_{\rm SS}$, which is identical to that in Ref. \cite{Badaracco2019}:
\begin{equation}
    C_{\rm SS}(\omega) = \mathcal{C}(\xi^{\rm tot}, \omega) \left[pf^{\rm P}(\Phi_{sk}^{\rm P})  + (1-p)f^{\rm S}(\Phi_{sk}^{\rm S})\right],
\end{equation}
where the indexes $s$ and $k$ represent two seismic sensors. Note that the $C_{\rm SS}(\omega)$ diagonal elements contain the seismometer's SNR (as explained in Reference \cite{Harms2019}, Equation (202)). The residual for the end TMs of the other two interferometers are calculated as described in Ref. \cite{Badaracco2019}.

\section{Results}\label{sec:results}
The Differential Evolution Optimizer, a stochastic algorithm, was chosen to find the minimum of the cost function. It was run 100 times for each array composed of $N$ sensors to select the best configuration among those selected by each single optimization. The results of these optimizations are discussed below. In the optimization the following distances from the origin were used for the TMs positions: $r_{in} = 64.12$\,m for the input TMs and $r_{end} = 536.35$\,m for the end TMs. The signal-to-noise ratio assumed for the seismometers was SNR = 15, the seismic velocities for S-waves and P-waves were respectively assumed to be: $v_S = 4$ and $v_P = 6$\, km/s and the frequency at which the array was optimized was 10\,Hz.

\section{Optimal arrays}
\begin{figure}[h!]
    \centering
     \includegraphics[width=0.85\textwidth]{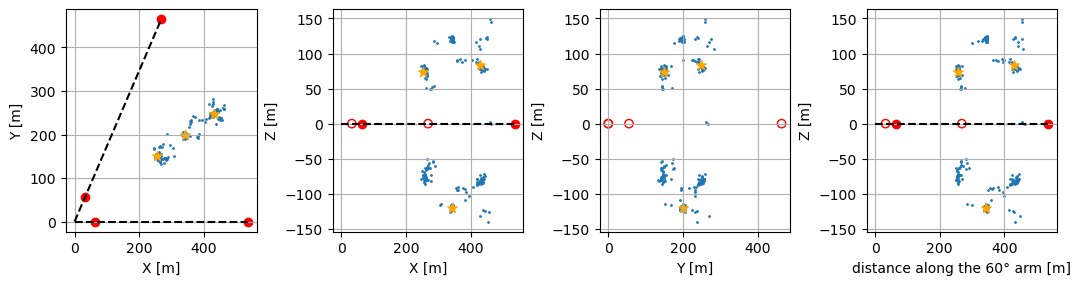}
     \includegraphics[width=0.85\textwidth]{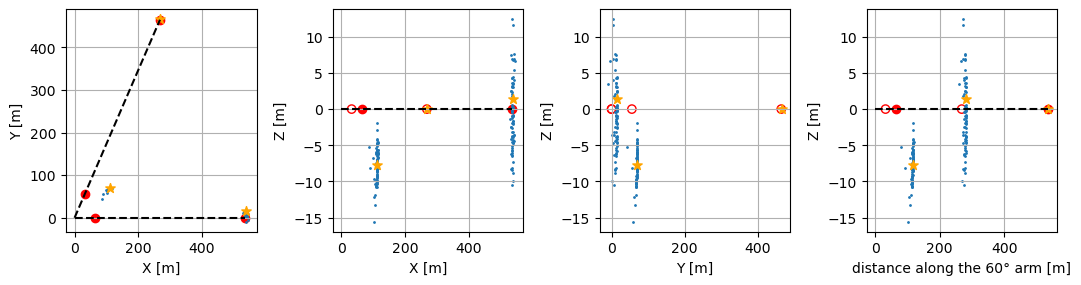}
     \includegraphics[width=0.85\textwidth]{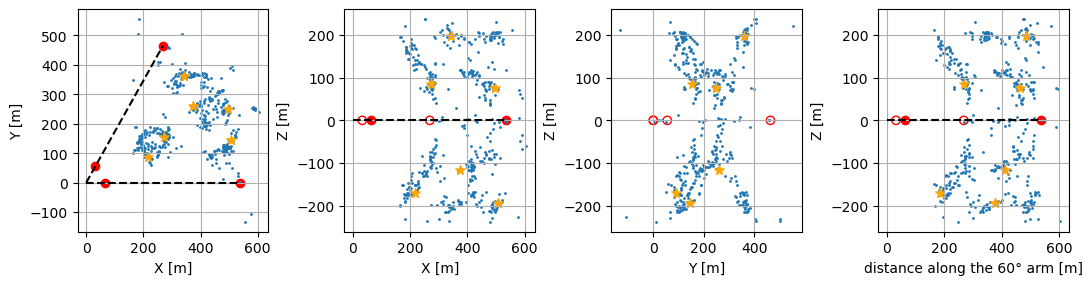}
     \includegraphics[width=0.85\textwidth]{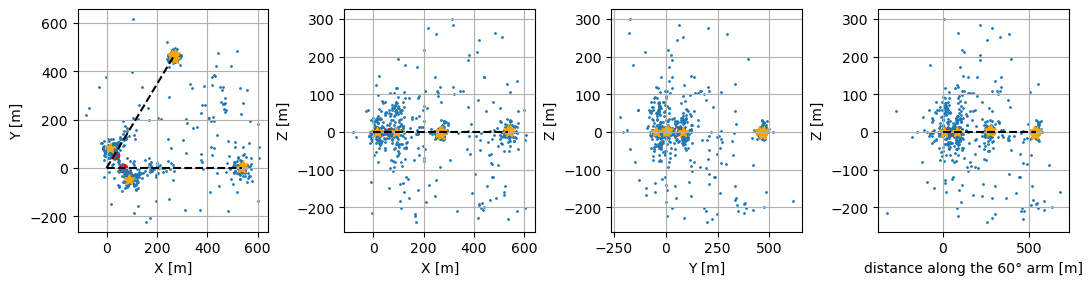}
     \includegraphics[width=0.85\textwidth]{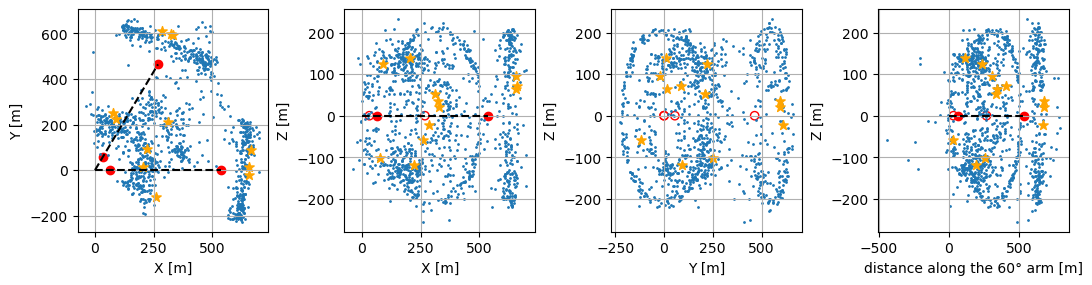}
    \caption{Projection of the optimal sensors (see text for more details). \textit{First row}: N=3 and $p$=0.2; \textit{Second row}: N=3 and $p$=0.1; \textit{Third row}: N=6 and $p$=0.2; \textit{Fourth row}: N=12 and $p$=0.0; \textit{First row}: N=12 and $p$=0.6;}\label{fig:optimal_arrays}
\end{figure}

In Fig.  \ref{fig:optimal_arrays}, the optimal arrays obtained from 100 optimizations are shown: the sensors belonging to the array producing the lower (i.e. the best) residual are indicated with yellow stars (the other sensors are represented as blue dots). Only the sensor arrays that generate a residual no greater than twice the variance calculated over all the 100 optimizations are displayed. It can be noticed that there is a certain degree of degeneracy: in some cases the sensors tends to be located around circles and/or arches. This is in agreement with the results of the previous work (Ref \cite{Badaracco2019}) and it is quite intuitive since we assumed an isotropic field. In other cases the sensors simply group around the same point or along straight lines, depending on the number of sensors used. This mostly occurs for extreme values of $p$ (0 or 1). When N grows, some sensors are located in the same place: this happens because placing more sensors reduces the SNR and a super-sensor is created.   

\subsection{Cancellation performances at different field compositions}
The value of $p$ in Eq. \ref{eq:p} represents the fraction of compressional waves in the seismic field. Values close to 1 indicate that the field is composed solely of compressional waves, while values close to 0 indicate that the field contains only shear waves. It is known that if both polarizations are present, NN cancellation capabilities of an array are reduced \cite{Badaracco2019}.

\begin{figure}[h!]
    \centering
    \includegraphics[width=1\textwidth]{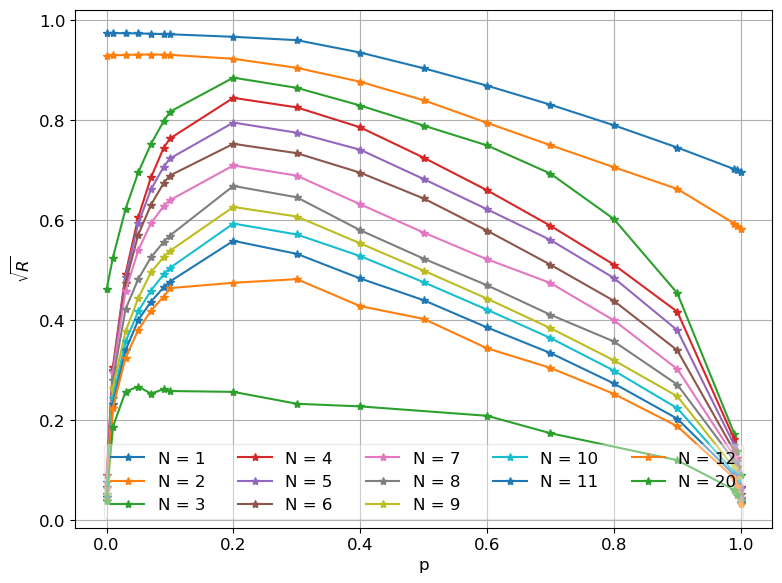}
    \caption{The plot shows how the residual changes varying the value of $p$ at a fixed number $N$ of sensors in the array. The instrument signal to noise ratio of the seismometers is assumed to be SNR = 15.}
    \label{fig:R_vs_p}
\end{figure}

In Fig. \ref{fig:R_vs_p}, we quantify how the performance (measured by the residual value) changes with respect to the field composition (value of $p$). We examined sensor arrays composed of $N$ sensors, with $N$ ranging from 1 to 15 and then 20. Fig. \ref{fig:R_vs_p} shows that NN cancellation performance strongly depends on the value of $p$, with the worst cases occurring for values between 0.1-0.4. The curve $R(p)$ is asymmetric due to the fact that compressional waves produce stronger NN (see Eq. \ref{eq:NN_acceleration}). This is the reason why at values close to 0 the residual worsens quickly: the array's ability to cancel shear-wave NN is compromised by the presence of compressional waves. Furthermore, for $N=1$ and 2 with $p$ closer to 1 or 0, cancellation performance is visibly worse than for other values of $N$. This is expected since the array is attempting to cancel the NN in three interferometers (three residual functions to be jointly minimized) using less than three seismometers. For 3 -- 12 seismometers, the shape of the curve depends weakly on the number of sensors, and as one would expect, the residual decreases with increasing number of sensors. 

Looking at Fig. \ref{fig:R_vs_p}, $p$ has an important impact on NNC performance only when a few sensors (per vertex) are deployed. Already with 20 seismometers, the value of $p$ does not have a strong influence unless it is close to its boundaries. However, it should be kept in mind that the value of $p$ has an impact on the optimal array configuration. 

In the NN literature, the value $p=1/3$ was typically used \cite{Harms2019, Badaracco2019,Amann2020, Janssens2022}. This assumption was made because the composition of the seismic was not known for sites of interest, and so the ad hoc assumption was made that the energy is equipartitioned among the three polarizations (one compressional mode and two shear modes). The actual value of $p$ will be different for each site and depend on the types of seismic sources and the local source distribution. An analytical model exists for a diffusive field produced by strong continuous or repeated scattering of waves from distant sources, which establishes a characteristic polarization mix $p$:
\begin{equation}\label{eq:weaver}
    p = \frac{1}{1+2(\alpha/\beta)^3}
\end{equation}
where $\alpha$ and $\beta$ represent the velocities of compressional and shear waves, respectively. This relationship was first derived by Weaver in 1982 \cite{Weaver82} and holds true in the diffuse regime independently of the details of the scattering processes. Equation \ref{eq:weaver} was derived for an infinite medium. Later, the model was extended to the case of a half-space taking into account the contribution of surface Rayleigh waves \cite{Perton2009}. The compressional-wave content of the diffusive field is small, but the model does not apply to our analysis where seismic waves in the NN band will not have passed through a lasting period of strong scattering to establish this polarization mix. The polarization mix of underground seismic fields in the NN band will strongly depend on the source properties and distribution. Since the sources of seismic waves at the ET candidate sites are mostly unknown and varying with time (some of them potentially located many kilometers away from the site), the only way to provide an informed estimate of $p$ is by analyzing data from seismic arrays.

\subsection{Robustness under small deviations from the optimal configuration}\label{sec:Robustness_only_displacement}
A NN cancellation system requires the deployment of a seismic array in optimal locations and the use of an optimal filter, such as the Wiener Filter, to process the data and cancel the noise. This necessitates two types of optimizations: one related to the spatial characteristics of the seismic field, providing the optimal positions of the array; and another related to the temporal characteristics of the seismic field, providing the estimates of the WF coefficients.

If the seismic field is non-stationary, it is possible to update the WF coefficients, to ensure that it remains the optimal filter over time provided that the field varies slowly enough to be able to adapt to the changes \cite{Tringali2019}. However, if the field is non-stationary, its spatial correlations can change as well affecting the optimal array positions. In an underground environment, it is not possible to update the optimal positions of the array. Spatial optimization must be a compromise between possible configurations of the seismic field. In this regard, it is preferable to have a site where seismic field correlations have minimal temporal variations.

\begin{figure}[h!]
    \centering
    \includegraphics[width=1\textwidth]{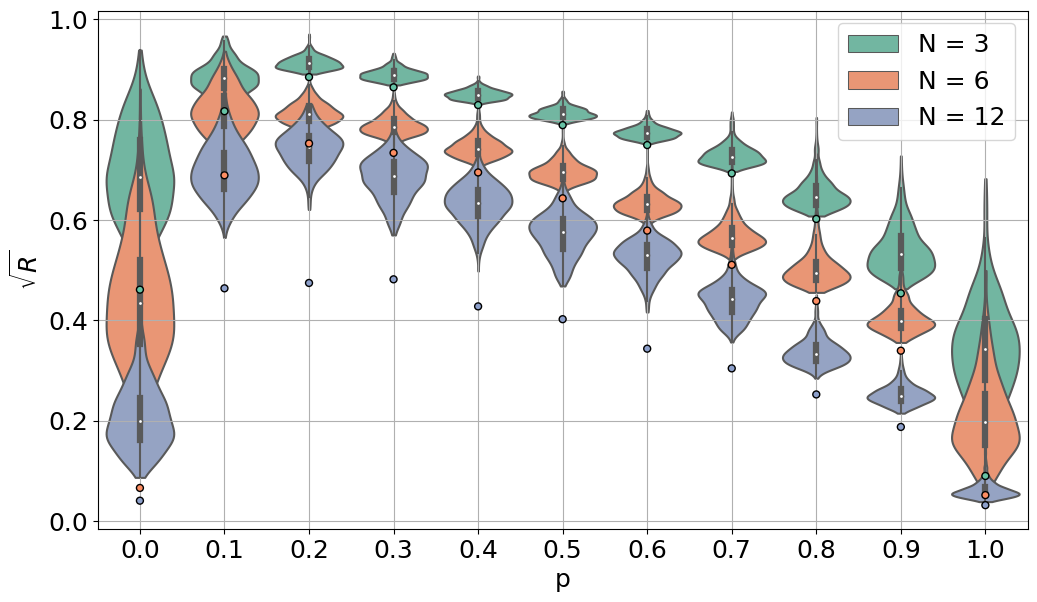}
    \caption{This figure illustrates the robustness of the array. For each value of $p$, the array was displaced from its optimal configuration and the residual was recalculated. The sensors were moved according to a normal distribution with a mean of 0\,m and $\sigma = 30\,$m. The coloured dots plotted over the violins indicate their respective best residual found with optimization. }\label{fig:robustness}
\end{figure}

Another important consideration is that the residual produced by an optimal seismic array will be higher than its nominal value, leading to reduced cancellation performance. This is because it will not be possible to deploy sensors in their exact optimal positions. In Fig. \ref{fig:robustness}, we tested the robustness of an optimal array for a homogeneous and isotropic seismic field to changes in sensor positions by randomly displacing sensors from their optimal positions. Fig. \ref{fig:robustness} shows how the residual changes: the coloured points represent the residual of the optimal array for a given $N$ (3, 6 and 13) and for each value of $p$. However, in a real-world scenario with seismic and geological inhomogeneities, robustness may differ, and whether seismometers can be deployed with a typical offset from optimal locations of 30\,m is unclear. Constraints on borehole positions due to surface conditions and infrastructure might well enforce larger offsets. Furthermore, since we run the optimization with the assumption of isotropy, we do not know yet by how much optimal configurations change under realistic time variations of seismic correlations.

\subsection{Robustness to a change in frequency with respect to the frequency used for the optimization}
\begin{figure}[h!]
    \centering
    \includegraphics[width=0.5\textwidth]{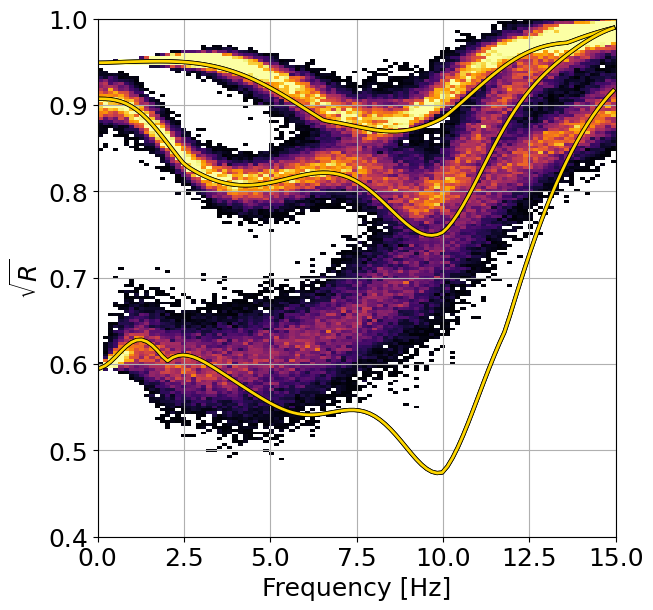}
    \includegraphics[width=0.5\textwidth]{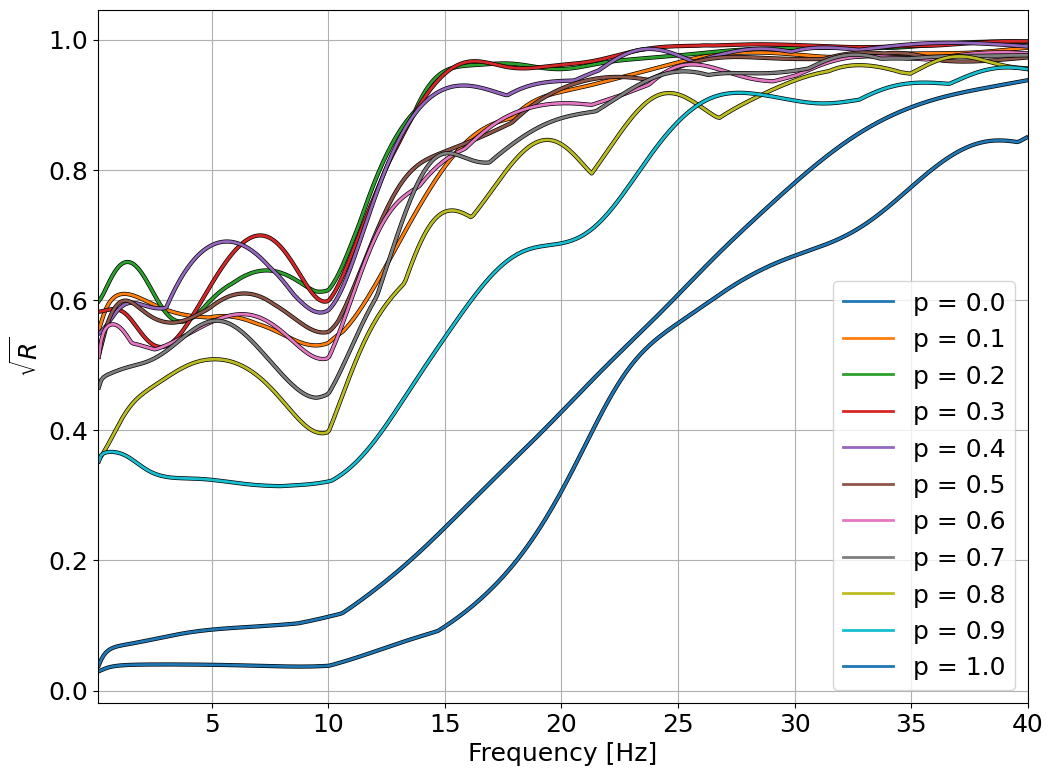}
    \caption{The \textit{upper} plot shows how the residual changes as the frequency varies at a fixed value of $p$ = 0.3 with $N=3$, 6, and 13 (from top to bottom). The colors represent the frequency of occurrence of the residual when the sensors are displaced from their optimal positions by a quantity drawn from a normal distribution with mean 0\,m and $\sigma = 30\,$m. The \textit{lower} plot shows how the residual changes as the frequency varies at different values of $p$ (with 10 sensors optimized at 10\,Hz.)}.
    \label{fig:R_vs_f_histo}
\end{figure}

\begin{figure} 
     \centering
     \includegraphics[width=\textwidth]{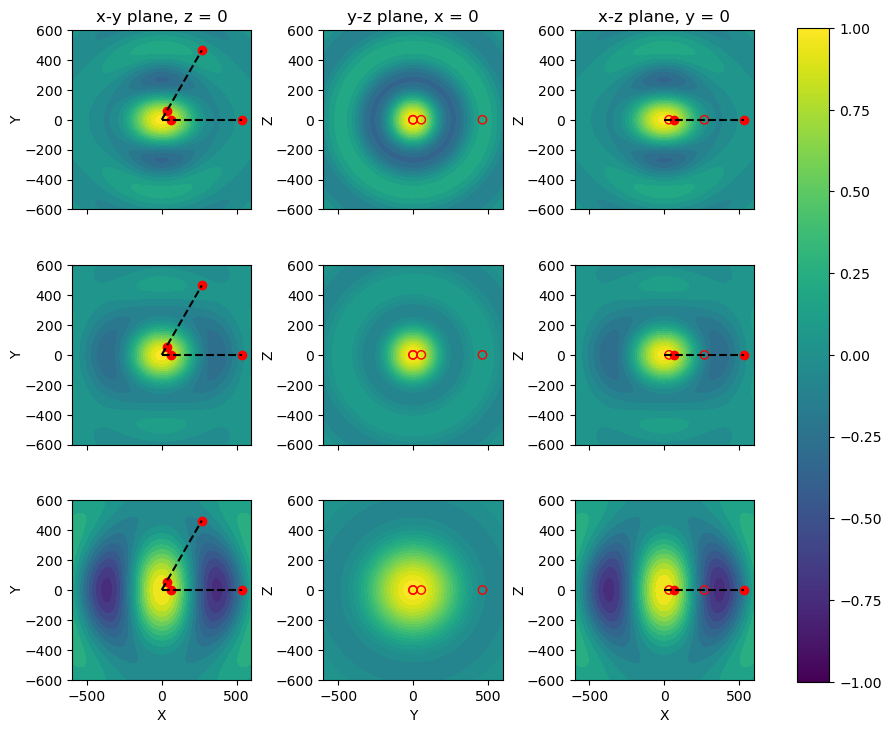}
     \caption{Seismic correlations between the origin point and all other points in the plane. The x-y, y-z and x-z planes are displayed. The ET layout is superimposed to provide a sense of distance. The full red circles represent the TMs laying on the plane, while the black dashed lines represent the ET baseline. The empty red dots represent the projection of the TMs not laying on the plane. Correlations are showed for a frequency of 10\,Hz and for $p$ = 0 (\textit{top} row), $p$ = 0.3 (\textit{middle} row) and  $p$ = 1.0 (\textit{bottom} row). }
     \hfill
     \label{fig:Corrp}
\end{figure}

The optimization is typically performed at a single frequency: the upper plot of Fig. \ref{fig:R_vs_f_histo} shows the residual calculated at other frequencies as a histogram indicating how the residual varies when the array is displaced from the optimal position as in the previous section. The optimization was done at 10\,Hz with $p$ = 0.3 for 3, 6 and 12 sensors. At frequencies higher than the optimal one, cancellation performances  worsen and above 15\,Hz cancellation becomes almost entirely ineffective. This is expected since the optimization was performed at another frequency. However, the behaviour at lower frequencies is opposite: the residual decreases even further for 13 sensors at 3\,Hz. This can be attributed to the fact that correlations increase at lower frequencies, meaning that the yellow area of Fig. \ref{fig:Corrp} becomes larger. Plotting the residual versus the frequency for a seismic array randomly positioned in the space surrounding an ET vertex reveals that the residual is close to $R=1$ at higher frequencies and reaches lower values below 15\,Hz. This effect is more emphasized when $p=0$ or 1, because correlations are larger even at higher frequencies. The same behaviour can be observed in the lower plot of Fig. \ref{fig:R_vs_f_histo}, where the residual of the optimal arrays at different values of $p$ are shown.

As a final note on broadband optimization, this type of optimization is particularly useful for expanding the frequency band where cancellation is most effective. However, it has the side effect of reducing overall performance compared to single-frequency optimization (see Ref. \cite{Badaracco2019}). One way to minimize this effect could be to check how the residual of an array optimized at a frequency $f_0$ changes at other frequencies. Examining the upper plot of Fig. \ref{fig:R_vs_f_histo}, it can be seen that there are some frequencies (other than the optimal one) where local minima occur. Therefore, broadband optimization could be performed at these specific frequencies (which should be larger than $f_0$ since the low-frequency residual is always good due to large correlations). For example, in the case $p=0.8$ of Fig. \ref{fig:R_vs_f_histo}, optimization could be done at 10, 16.1 and 21.2\,Hz where local minima are clearly visible.    


\section{Conclusions}
This paper builds upon previous work reported in Ref. \cite{Badaracco2019} by taking into account the triangular shape of ET to optimize the seismometer array for effective Newtonian-noise cancellation at an entire vertex of the detector. We studied the impact of the seismic field composition on noise cancellation in terms of body-wave polarizations and the robustness of the performance with respect to changes in optimal frequency and positions.

The triangular shape of ET entails that the input test masses of an interferometer will be close to the end test masses of two other interferometers. Correlations of Newtonian noise between test masses and seismic correlations were taken into account in our study (see also Ref. \cite{Janssens2022} for a study on the impact of correlated noise in ET). The optimal arrays calculated in this study minimized the residuals simultaneously in three interferometers.

We find that under the assumption of an isotropic, stationary field, a factor 5 reduction of Newtonian noise in amplitude can be achieved with 20 seismometers provided that they are located precisely at their ideal positions. A deviation from the optimal positions degrades the performance depending on the field composition and number of sensors. For example, when the sensors of an array with 12 seismometers are randomly displaced from their ideal positions by (typically) 30\,m, the residuals are larger by (in average) a factor 1.45 in amplitude, and performance degrades more strongly when either compressional or shear waves dominate instead of having a more uniform polarization mix. 

While this study is based on idealized conditions (isotropy, stationarity), which will not be found at the ET candidate sites, it provides important constraints on what can realistically be achieved with Newtonian-noise cancellation. The optimal placement of seismometers in boreholes will be a formidable challenge since our understanding of seismic correlations underground are very limited, and correlations can vary with time. The work presented here now needs to be connected to more realistic models of the seismic field to estimate more accurately how good and robust the performance of the ET Newtonian-noise cancellation system will be.

\section{Acknowledgements}

This article is based upon work from COST Action CA17137, supported by COST (European Cooperation in Science and Technology). FB and LR thank the experiment next\_AIM which provided computational resources for the analysis. F.B would like to express her deepest appreciation to Francisco Sánchez-Sesma for the time he dedicate in providing his invaluable assistance in finding and understanding the correct bibliography on the theory of equipartitioning of diffuse seismic fields. F.B would like also to extend her thanks to Matteo Di Giovanni who provided help in understanding the correct values for seismic velocities at Sos Enattos and Luca Naticchioni for providing valuable insights on positioning errors that can be made while excavating boreholes.  Finally, a thanks goes also to Giovanni Luca Cardello who provided clarifications about the seismology and geology of Sos Enattos.

\bibliographystyle{plain}
\bibliography{Bibliography}
\end{document}